\documentclass[aps,prd,showpacs,preprintnumbers,twocolumn,groupedaddress,nofootinbib]{revtex4}
\usepackage{graphicx}
\usepackage{latexsym}
\usepackage{booktabs}
\usepackage{multirow}
\usepackage{amsmath}
\usepackage{color}
\usepackage[usenames,dvipsnames]{xcolor}
\pdfoutput=1

\def\beq{\begin{equation}}
\def\eeq{\end{equation}}
\def\bey{\begin{eqnarray}}
\def\eey{\end{eqnarray}}

\def\lsim{\mathrel{\raise.3ex\hbox{$<$\kern-.75em\lower1ex\hbox{$\sim$}}}}
\def\gsim{\mathrel{\raise.3ex\hbox{$>$\kern-.75em\lower1ex\hbox{$\sim$}}}}

\newcommand\CW[1]{{#1}}
\newcommand\EC[1]{{#1}}

\begin{document}

\title{A Clustering Analysis of the Morphology of the 130 GeV Gamma-Ray
\CW{Feature}}
\author{Eric Carlson$^{1}$, Tim Linden$^{1}$, Stefano Profumo$^{1,2}$ and
Christoph Weniger$^{3}$}
\affiliation{$^1$ Department of Physics, University of California, Santa Cruz,
1156 High Street, Santa Cruz, CA, 95064}
\affiliation{$^2$ Santa Cruz Institute for Particle Physics, University of
California, Santa Cruz, 1156 High Street, Santa Cruz, CA, 95064}
\affiliation{$^3$ GRAPPA Institute, University of Amsterdam, Science Park 904,
1090 GL Amsterdam, Netherlands}

\begin{abstract}
\noindent Recent observations indicating the existence of a monochromatic
$\gamma$-ray line with energy $\sim$130~GeV in the Fermi-LAT data have
attracted great interest due to the possibility that the line feature stems
from the annihilation of dark matter particles.  Many studies examining the
robustness of the putative line-signal have concentrated on its spectral
attributes.  Here, we study the morphological features of the
$\gamma$-ray line photons, which can be used to differentiate a putative dark
matter signal from astrophysical backgrounds or instrumental artifacts.
Photons stemming from dark matter annihilation will produce events tracing a
specific morphology, with a statistical clustering that can be calculated
based on models of the dark matter density profile in the inner Galaxy. We
apply the DBSCAN clustering algorithm to Fermi $\gamma$-ray data, and show
that we can rule out the possibility that 1 (2, 4) or fewer point-like
sources produce the observed morphology for the line photons at a $99\%~(95\%, 90\%)$  confidence level. Our study
strongly disfavors the main astrophysical background envisioned to produce a
line feature at energies above 100 GeV: cold pulsar winds. It is highly unlikely
that 4 or more such objects have exactly the same monochromatic cosmic-ray
energy needed to produce a $\gamma$-ray line, to within instrumental energy
resolution.  Furthermore, we show that the larger photon statistics expected
with Air Cherenkov Telescopes such as H.E.S.S.-II will allow for
extraordinarily stringent morphological tests of the origin of the ``line
photons''.
\end{abstract}
\pacs{98.70.Rz, 95.55.Ka, 95.35.+d, 97.60.Gb}

\maketitle

\section{Introduction}
\label{sec:introduction}

The launch of the Fermi Large Area Telescope (LAT) in 2008 has allowed for a
greatly expanded view of the $\gamma$-ray sky, including a significantly
enhanced energy and angular resolution, compared to previous
missions~\citep{atwood_fermi_mission}. These characteristics have allowed, in
particular, for a thorough investigation of the extremely dense population of
high-energy $\gamma$-ray sources in the Galactic center (GC)
region~\citep{gc_population_of_sources}. The GC region is known to
host such diverse $\gamma$-ray sources as supernova
remnants~\citep{gc_snr}, highly ionized gas~\citep{gc_highly_ionized_gas},
dense molecular clouds~\citep{ferriere_dense_molecular_clouds}, massive O/B
stars~\citep{gc_stellar_population}, both young and recycled pulsar
populations~\citep{gc_possible_pulsar_populations}, as well as being the
densest region of dark matter in the Galaxy~\citep{gc_dark_matter}. Notably,
no other location in the sky is expected to provide a signal from dark matter
annihilation which is as bright as the GC (for a recent general review of
gamma-ray searches for signals from dark matter annihilation see
Ref.~\cite{Bringmann:2012ez}).  While this makes the GC region an extremely
interesting location for a multitude of scientific studies, it also means that additional information, such as characteristic spectra
for each source class, must be carefully considered in order to separate the
desired signal from the bright background.

Recently, \citet{bringmann_internal_bremsstrahlung} and \citet{weniger_line}
found indications for a unique spectral signature in observations of the region surrounding the
GC, which is consistent with dark matter annihilation. Specifically, in sky regions optimized for large signal-to-noise ratios for various
dark matter density profiles, they observed an excess of photons with an
energy spectrum resembling a 130~GeV $\gamma$-ray line, smeared by the finite
energy resolution of the Fermi-LAT telescope. The significance of this excess
was found to be $3.2\sigma$ globally~\cite{weniger_line} -- enough to make the
feature interesting on statistical grounds. The feature is strongest when
using regions of interest that have been optimized for dark matter density distributions following: (1) a Navarro-Frenk-White (NFW)  profile~\citep{nfw}, (2) an Einasto profile~\citep{einasto}, and (3) a generalized NFW profile, with a  radial slope governing the dark matter density profile (r$^{-\alpha}$) set to $\alpha=1.15$, similar to what could
result from adiabatic contraction \citep{blumenthal_adiabatic_contraction,
gnedin_adiabatic_contraction}. The reported feature is much weaker for profiles where the dark matter density is cored near the GC.

A monochromatic $\gamma$-ray line has been long considered the ``Holy Grail''
for dark matter indirect detection, given the difficulty of producing a
high-energy monochromatic signal with ordinary astrophysical processes. Thus,
this observation prompted a number of follow-ups.
\citet{linden_profumo_bubble} noted that the observation of a monochromatic
$\gamma$-ray signal could be qualitatively mimicked by an
additional power-law component which breaks strongly at an energy of 130~GeV,
and posited the Fermi bubbles~\citep{su_slatyer_finkbeiner_bubbles} as a
possible source for this excess (although rather strong breaks are required to
fit the data, see Ref.~\cite{Bringmann:2012ez}). Most notably,
\citet{su_finkbeiner_line} localized the emission to within approximately
5$^\circ$ of the GC, finding a 6.5$\sigma$ (5.0$\sigma$ after including a
trials factor) preference for a line signal following an off-center Einasto
profile compared to the assumption of a simple power-law background (see also
\citep{tempel_line_isolation}). This finding disputes the implication of the
Fermi bubbles as a source for the excess, as the latter are observed to extend
over a much larger emission region. The off-centered nature of the dark matter
profile is at the moment only marginally statistically
significant~\citep{line_offset_from_gc_is_statistical, Rao:2012fh}. However,
if confirmed, an off-center peak would pose a challenge to
traditional models of the Galactic dark matter density which assume the largest dark matter density to fall on top of the peak of the baryonic mass density\footnote{Which we assume,
  here, to be identical to the position of Sgr A*, and call the GC throughout
this paper.}. However, a recent analysis by \citet{kuhlen_peak_of_dm_density} found that
the peak in the dark matter density may, in fact, be displaced by hundreds of
parsecs from the dynamical center of the Galaxy - although this scenario might
be incompatible with the assumption of a cusped profile.

More data are needed to clarify whether the 130 GeV feature in the data
persists with larger statistics or if it is only a statistical fluke. Since
the observation of a $\gamma$-ray line is extremely sensitive to inaccuracies
in the energy-reconstruction of $\gamma$-rays observed by Fermi-LAT, a great
deal of interest has also focused on searching for possible instrumental
abnormalities affecting the photons belonging to the $\gamma$-ray line. The most
notable characteristic of any such instrumental effect would be the
observation of line activity either across the entire sky, or across a certain
region of instrumental phase space.  Interestingly, some early results found
an excess of 130~GeV events in observations of the
Earth-limb~\cite{su_finkbeiner_line, finkbeiner_systematics, Hektor:2012ev}.
This is troubling, as the vast majority of limb photons are known to result
from the di-photon $\pi^0$ decay spectrum created as cosmic-ray protons
interact with the upper layers of the Earth's atmosphere. There is no
conceivable model in which dark matter annihilation could create a
$\gamma$-ray line in this region~\citep{su_finkbeiner_line}. However, this
line activity was not
detected along the Galactic plane, which provides significantly more photons
than the GC, and inhabits similar regions of instrumental phase space. A
comprehensive study by \citet{finkbeiner_systematics} did not find any
significant evidence for systematic features in the energy reconstruction of
the Fermi-LAT, which would be able to artificially produce a $\gamma$-ray line
(see however Refs.~\cite{Boyarsky:2012ca, Whiteson:2012hr, Whiteson:2013cs}). 

Follow up analyses by the Fermi-LAT team are currently ongoing, but have
revealed two noteworthy results. At the time of the initial discovery of the
$\gamma$-ray line, efforts were already ongoing to improve the energy
normalization of the Fermi-LAT data, accounting for a decrease in the
calculated energy of $\gamma$-rays over time due to radiation damage to the
calorimeter. This effect linearly increased the reconstructed energy of high
energy photons, moving the line signal from 130 GeV up to approximately 135
GeV. However, this reprocessing did not greatly affect any other signature of
the posited line analysis. We note that throughout the rest of this paper, we
refer to the ``130~GeV line", as we are using a version of the Fermi-LAT data
which has not been reprocessed. However, all results shown here are very
nearly applicable to an analysis of the 135~GeV line observed in the
reprocessed data. 

Additionally, the Fermi-LAT analysis did uncover one troubling aspect of the
posited $\gamma$-ray line. Employing a parameter CTBBestEnergyProb (which is
not publicly available), they investigated the confidence they had in the
energy reconstruction of each photon belonging to the $\gamma$-ray line. In
the case where a true $\gamma$-ray line feature is present in the data, this
should increase the statistical significance of the observation, as the
proceedure adds additional statistical weight to the line photons which are
most likely to have a correctly measured energy. However, when this analysis
was applied to the observed photon data, the statistical significance of the
line feature was found to decrease moderately. This signals that the line
feature has photons with a somewhat poorer energy resolution than would
generally be expected\citep{bloom_charles_fermi_lat_line}. However, further
inquiry of these systematic issues is required, as none of the systematics can
clearly account for the entire statistical strength of the line feature.

If interpreted as a signal of particle dark matter annihilation or decay, the
large observed intensity of the 130~GeV $\gamma$-ray line (along with strong
constraints on the total continuum emission from additional hadronic
states~\citep{hooper_linden_gc}) has proved a
difficult, though by no means intractable, particle physics problem. Numerous
models have already been posited to ``brighten" the $\gamma$-ray
line~\citep{line_pp_1, line_pp_2, line_pp_3, line_pp_4, line_pp_5, line_pp_6,
line_pp_7, line_pp_8, line_pp_9, line_pp_10, line_pp_11, line_pp_12,
line_pp_13, line_pp_14, line_pp_15, line_pp_16, line_pp_17, line_pp_18,
line_pp_19, line_pp_20, line_pp_21, line_pp_22, line_pp_23, line_pp_24,
line_pp_25, line_pp_26, line_pp_27, line_pp_28, line_pp_29, Asano:2012zv}.
Summarizing the myriad particle physics details of these models lies beyond
the scope of the present paper. 

Although often characterized as a ``smoking gun'' signature for the
annihilation or decay of particle dark matter, tentative observations of the
130 GeV line have spurred the question of whether any traditional
astrophysical mechanism might mimic a line in the relevant energy
range. \citet{aharonian_cold_pulsar_winds} argued that the only plausible
mechanism for the creation of an astrophysical $\gamma$-ray line is through
inverse Compton scattering of ambient photons by a jet of nearly monoenergetic
electrons and/or positrons, occurring in the deep Klein-Nishina regime. If the
latter kinematic regime
holds, the photon acquires nearly the entire energy of the incoming lepton,
allowing for a nearly monoenergetic lepton spectrum to efficiently transfer
into a sharply peaked $\gamma$-ray feature.

One possible class of astrophysical objects that possesses the potential to
host the needed leptonic monochromatic ``jet'', as well as the ambient photons
in the needed energy range (here, a few eV) is {\em cold ultrarelativistic
pulsar} (PSR) {\em winds} \cite{aharonian_cold_pulsar_winds}. It is important
to note that this scenario presents a potential difficulty in explaining the
observed spread of 130~GeV photons beyond a single point source, as different
PSRs would be expected to exhibit $\gamma$-ray lines at different energies.
One way to test this one astrophysical background is therefore to study
whether the morphology of the observed 130 GeV photons can be reproduced with
a small number of point sources or not. This is the key objective of the
present study.

Several other approaches have tested the dark matter nature of the
130 GeV photons. For example, any dark matter interpretation of the 130~GeV
line implies additional regions of interest for follow up searches, where the
so-called $J$-factor (the line-of-sight integral of the dark matter density
squared, smeared over the instrumental point-spread function) is expected to
be largest. Most importantly, dwarf spheroidal galaxies and galaxy clusters
have been singled out as promising regions to search for a dark matter signal.
Observations of Milky Way dwarfs have not uncovered any evidence of a 130~GeV
signal~\citep{dwarf_spheroidal_130_GeV_line}: this is not unexpected, as the
estimated annihilation cross-section to $\gamma \gamma$ implied from GC
observations predicts less than one photon to arrive from the population of
dwarf spheroidal galaxies. Interestingly, \citet{galaxy_cluster_130_GeV_line}
argued for an observation of a 130~GeV line in a population of nearby galaxy
clusters. However, it should be noted that this signature is only significant
when very large ROIs of $\sim$8$^\circ$ are considered, which is much larger than
the expected angular size of the galaxy clusters under investigation.

Using a similar method, \citet{su_finkbeiner_line_unassociated} investigated
the population of unassociated Fermi-LAT point sources -- that is, point
sources detected by the Fermi-LAT instrument which have not been identified at
other wavelengths. They found a statistically significant detection for a
double 130~GeV and 111~GeV line, with 14 unassociated sources showing evidence
of a line photon. Furthermore, they found no significant detection of
$\gamma$-ray line emission in the control sample of Fermi-LAT point sources
that have already been associated with various astrophysical phenomena. This
caused them to conclude that some portion of the unassociated point-source
sample may
contain previously unknown dark matter substructures. However,
\citet{hooper_linden_unassociated} argue against this conclusion, noting that
each unassociated source is identified primarily based on its continuum
emission between energies of 100~MeV-10~GeV, rather than based on the
detection of a single line photon.  The intensity and spectrum of this
continuum emission can then be compared to the signal from dark matter
annihilation to any final state producing a $\gamma$-ray continuum, and is
expected in order to produce the thermal relic abundance of dark matter. They
find that for at least 12 of the 14 indicated unassociated point sources, the
continuum emission is not compatible with any dark matter annihilation
pathway. Furthermore, they argue that the latitude distribution of the
identified sources is not consistent with that expected from any model of dark
matter subhalo formation. A second analysis by
\citet{mirabal_unassociated_are_agns} argued that while these 14 sources
remain unidentified, at least 12 of the sources (not identical to those from
\citep{hooper_linden_unassociated}) are spectrally  strongly consistent with
AGNs.

In addition to considering the energy signature of the 130~GeV line,
understanding the morphology of the photons belonging to the source
class producing the observed feature will be key to elucidate the physics
behind the line phenomenon. Notably, the point-spread function of
front-converting events at energies near 100~GeV approaches
0.1$^\circ$~\citep{atwood_fermi_mission}, which is significantly smaller than
the $\sim$5$^\circ$ region of interest implicated by
~\citep{su_finkbeiner_line}, allowing for the actual morphology of the line
emission to be closely mirrored by Fermi-LAT observations. To first order,
observations indicating a morphology consistent with widely accepted dark
matter density profiles would provide additional evidence for a dark matter
interpretation (the data indeed points to that direction, see e.g. Fig.~3 of
Ref.~\cite{Bringmann:2012ez}), while measurements consistent with either a
population of point sources, or a significantly flattened profile may point to
other astrophysical or instrumental interpretations.

In this \emph{paper}, we examine the morphology of photons belonging to the
$\gamma$-ray line more closely, analyzing with a sound statistical approach
the distribution of the arrival direction of photons by employing a clustering
algorithm to pinpoint the correlation between the arrival directions of
photons putatively belonging to the line feature. We then compare these
results against simulated models where the line is produced by
dark matter annihilation or by an astrophysical process associated with a few
point sources (for example a handful of PSRs). The key result of our study is
that {\em current data disfavor a scenario where the line photons stem from
4 or fewer point sources}. Given how unlikely it is that 4 or more
pulsars produce a gamma-ray line at exactly the same energy (within the LAT
energy resolution), our study disfavors the PSR scenario over a truly diffuse
and un-clustered origin for the photons.

In Section~\ref{sec:models} we describe the data employed in our observations
of the $\gamma$-ray line feature, the specifics of the algorithms used to
determine the photon morphology, our models for both the annihilation of
dark matter and emission from PSRs, and the diffuse background in the
region. In Section~\ref{sec:results} we present the results of our study both
for current Fermi observations, and for projections for upcoming observations
of the GC with the Atmospheric Cherenkov Telescope (ACT) H.E.S.S.-II.
Finally, in Section~\ref{sec:conclusions} we discuss the interpretation of the
results, and present our conclusions.

\section{Models}
\label{sec:models}

\subsection{Photon Selection}
In order to analyze the population of photons stemming from the putative
$\gamma$-ray line emission, we must make a photon selection which isolates the
line photons from those correlating to background events. We follow here the
same photon selection employed in Ref.~\cite{weniger_line_observations}, which
provides the location of observed Fermi-LAT photons in three energy bands,
70-110~GeV, 120-140~GeV, and 150-300~GeV over a 10$^\circ$ square window
centered on the GC.  Since the Fermi-LAT energy resolution is approximately
10\% at 130~GeV, we assume photons in the  120-140~GeV band to encompass the
photons related to the $\gamma$-ray line observation, while photons in the low
and high energy bands correspond to background events not associated with the
$\gamma$-ray line observation. Below, a power-law fit to the sidebands will be used to fix the
background rate in our 120-140 GeV simulations while the remainder of the photon excess will make up the signal.  We note that there is some evidence for a second
line at energies of around 111~GeV~\citep{su_finkbeiner_line}, however the
weak significance of that feature makes its impact on the population of
70-110~GeV $\gamma$-rays negligible. 

In comparing the photon morphology from the line region against the
``side-band'' photons, we assume that the background morphology remains
approximately invariant throughout the 70-300~GeV energy range. This
assumption is warranted in light of observations indicating that the primary
component of diffuse emission through this region stems from $\pi^0$-emission
tracing the Galactic gas~\citep{tev_diffuse_emission_plane, hooper_linden_gc}.
While some unresolved point-sources may also be present, it is unlikely that
any given source contributes multiple photons to the observed high-energy
$\gamma$-ray emission, making the spectral features of each individual source
irrelevant.

Before proceeding with the description of the clustering algorithm we employ
in this analysis, we describe in the following sections the simulated data
sets we use to validate our analysis. Section \ref{sec:dmsrc} details the
simulated line events from both dark matter annihilation scenarios with different dark matter density profiles, and scenarios with one or more point
sources. Section \ref{subsec:bg_model} details the simulated
background events. Finally, sec.~\ref{sec:clustering} describes the clustering
algorithm we employ in the present study.

\subsection{Dark Matter and Pulsars Models}\label{sec:dmsrc}
In order to establish a quantitative measure for the clustering properties of
$\gamma$-rays due to either dark matter or one or more point-sources in the GC
region, we produce Monte Carlo simulations of the expected positions of photons
stemming from each model.

In the case of PSRs,  we
examine models featuring between 1 and 6 point sources to explain the excess
130~GeV emission. We randomly pick the distribution of each point source
following a surface density distribution $\rho$(r)~$\propto$~r$^{-1.2}$, as
motivated by the observed density distribution of O/B stars in the inner
Galaxy~\citep{catchpole_surface_density_bright_stars} and we produce an excess number of photons which are distributed randomly (assuming equal brightness) between the simulated pulsars.

In the case of dark matter annihilation, we predict the annihilation signal to
follow the integral over the line of sight of the square of the dark matter
density. We choose two independent dark matter density profiles motivated by
models of dark matter structure formation. We first examine a generalized
Navarro-Frenk-White~\citep{nfw} profile, with a density profile

\begin{equation}
  \label{eq:nfw}
  \rho(r) \propto \Bigg(\frac{r}{r_s}\Bigg)^{-\alpha}\Bigg(
  1+\frac{r}{r_s}\Bigg)^{-3+\alpha}\;.
\end{equation}
In our standard analysis we choose $\alpha$~=~1 and r$_s$~=~22~kpc fitting the
best numerical results from the Aquarius simulation~\citep{einasto}. In order
to evaluate the effect of changing the dark matter density profile, we also
consider an Einasto profile with a density distribution~\citep{einasto}:

\begin{equation}
  \label{eq:einasto}
  \rho(r) \propto
  \exp\Bigg[-\frac{2}{\alpha}\Bigg(\Bigg(\frac{r}{r_s}\Bigg)^{-\alpha}
  -1\Bigg)\Bigg]\;,
\end{equation}
assuming, here, that $\alpha$~=~0.17~\citep{einasto}. In each case, we assume
that the annihilation rate is proportional to $\rho^2$(r), and then integrate
over the line of sight from the solar position R$_\odot$~=~8.3~kpc
\citep{solar_position} in order to generate the dark matter morphology that
would be observed by the Fermi-LAT. 

We additionally consider two alternative profiles. First, the case of decaying
dark matter following the NFW profile given in Eq.~\eqref{eq:nfw}, with the
decay rate now proportional to $\rho$(r).  Second, the case of
isotropic emission, i.e. a uniform surface profile. Since the clustering
properties of the source are highly dependent on the number of observed
photons, we calculate for Fermi-LAT (H.E.S.S.-II), $10^5$ ($2000$)
realizations of 48 (5000) photons following the distribution assumed in each
of these cases over a $10^\circ$ ($4^\circ$) square window. For H.E.S.S.-II observations, we estimate the
number of photons from a relatively short exposure time, on the order of a 6.25
hours, using an effective area given for the H.E.S.S.-II telescope with the flux in the 130~GeV energy range measured by the Fermi-LAT.

In each case, we must also consider the smearing of target photons based on
the point-spread function of the Fermi-LAT telescope. In order to accomplish
this accurately for observations at the GC, we employ the Fermi tools to
estimate the point-spread function for photons entering both the front and
back of the instrument at different $\theta$-angles.  Specifically, we employ
the \emph{gtpsf} tool developed by the Fermi-LAT collaboration in order to
calculate the effective PSF given the total exposure of the GC region from all
locations in the Fermi-LAT instrumental phase space (i.e. how much was the GC
viewed from different spacecraft orientations).  For the selected observation
period, the average (68\%, 95\%) containment radius over the
observation is (0.124$^\circ$,0.529$^\circ$) for front-converting events ($\sim56\%$ of
exposure area) and (0.258$^\circ$,0.907$^\circ$) for rear converting events ($\sim44\%$ of
exposure area).  The resulting PSF for each photon (signal and background) is
randomly chosen based on this weighted average of instrument coordinates and
the incoming photon is smeared based on the given PSF.  In the case of
Atmospheric Cherenkov Telescope (ACT) simulations, the PSF depends somewhat
sensitively on the angle of incidence of the incoming photons.  Following Aharonian \emph{et al} \cite{HESS_PSF} 
we approximate the H.E.S.S. point-spread function as an energy independent, two-component Gaussian with the probability density of an event smearing to radius $\theta$ given by, 

\begin{equation}
 P(\theta) = A \theta \left[\exp \left(-\frac{\theta^2}{2\sigma_1}\right) + A_{rel}\exp \left(-\frac{\theta^2}{2\sigma_2}\right) \right]
\end{equation}

Where $\sigma_1=0.046$, $\sigma_2=0.12$, and $A_{rel}=0.15$ and an overall normalization $A$.

\subsection{Background Models}
\label{subsec:bg_model}
In order to characterize the morphology of the expected diffuse background we
follow the detailed PASS 7 Galactic Diffuse Model, which contains both
spectral and morphological information generated by observations of both HI
and CO line surveys, which constrain the distribution of interstellar gas.
The $\gamma$-ray morphology and spectrum are then generated by convolving
these maps with the modeled cosmic-ray densities utilizing the Galprop code
\cite{galprop}, and calculating the expected $\gamma$-ray emission from processes
including $\pi^0$-decay, bremsstrahlung emission, and inverse-Compton
scattering.  Utilizing these simulations, we then generate a Monte Carlo
population of background $\gamma$-rays following a morphology compatible with
observations across the $\gamma$-ray spectrum.  

In the case of the Fermi-LAT telescope, we assume zero cosmic-ray
contamination, since we are considering only the GC region, which is very
bright in $\gamma$-rays. Given the calculated intensity of the $\gamma$-ray
line, we calculate an average of 12 signal, and 36 background photons between
an energy of 120--140 GeV. In each simulation of the Fermi-LAT data, we allow
the strength of this signal to float using Poisson statistics, setting the
mean intensity to be 12 signal photons. For Fermi-LAT observations we model a
10$^\circ$ square window around the GC. 

In the case of ACT observations, we note that cosmic-ray contamination is a
much larger issue, since hadronic showers dominate the data collected by these
telescopes. Extrapolating the cosmic-ray and $\gamma$-ray signals from recent
low-energy H.E.S.S. observations at 300~GeV~\citep{2011_HESS} and using the
best estimates for the H.E.S.S.-II instrumental
characteristics~\citep{Bringmann:2011ye}, we find that 86\% of the total
background signal will stem from cosmic-ray backgrounds. Thus, for ACT
observations we create a simulation composed of 4.35\% signal photons, 13.15\%
diffuse background photons (following the Fermi PASS-7 Galactic diffuse model)
and 82.5\% isotropic background photons. As in the case of Fermi-LAT
observations, we allow the total number of signal photons to float using
Poisson statistics, and maintain a background which is 86\% isotropic, and
14\% diffuse. For H.E.S.S.-II observations we model a 4$^\circ$ square window
around the GC, consistent with the smaller field of view of ACT instruments.

The background model described above is also used to estimate the average background
count $N_b$ during the computation of the cluster significances. $N_b$ is
calculated by integrating the background template, normalized to the correct
background count over the 95\% containment area of the cluster members (100\%
containment in cases with fewer than 20 cluster members).  This allows for a
statistical measure which traces the local morphology of the background and is
thus minimally dependent on its anisotropic structure, reducing the
significance of `hot-spots' in the background which may be falsely identified
as true clusters.

\begin{figure*}[htp]
  \begin{tabular}{cc}
    \includegraphics[width=.45\linewidth]{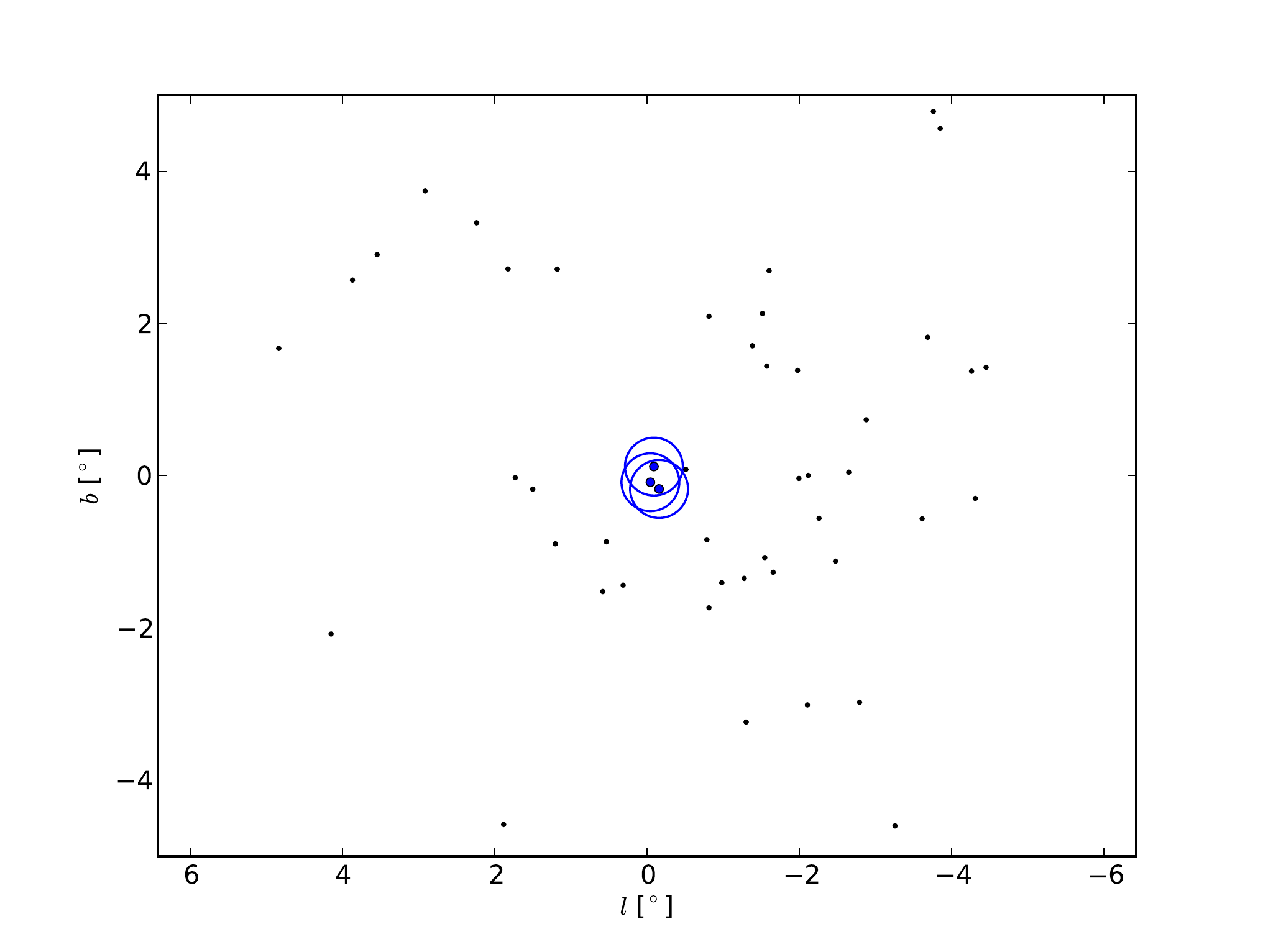} &
    \includegraphics[width=.45\linewidth]{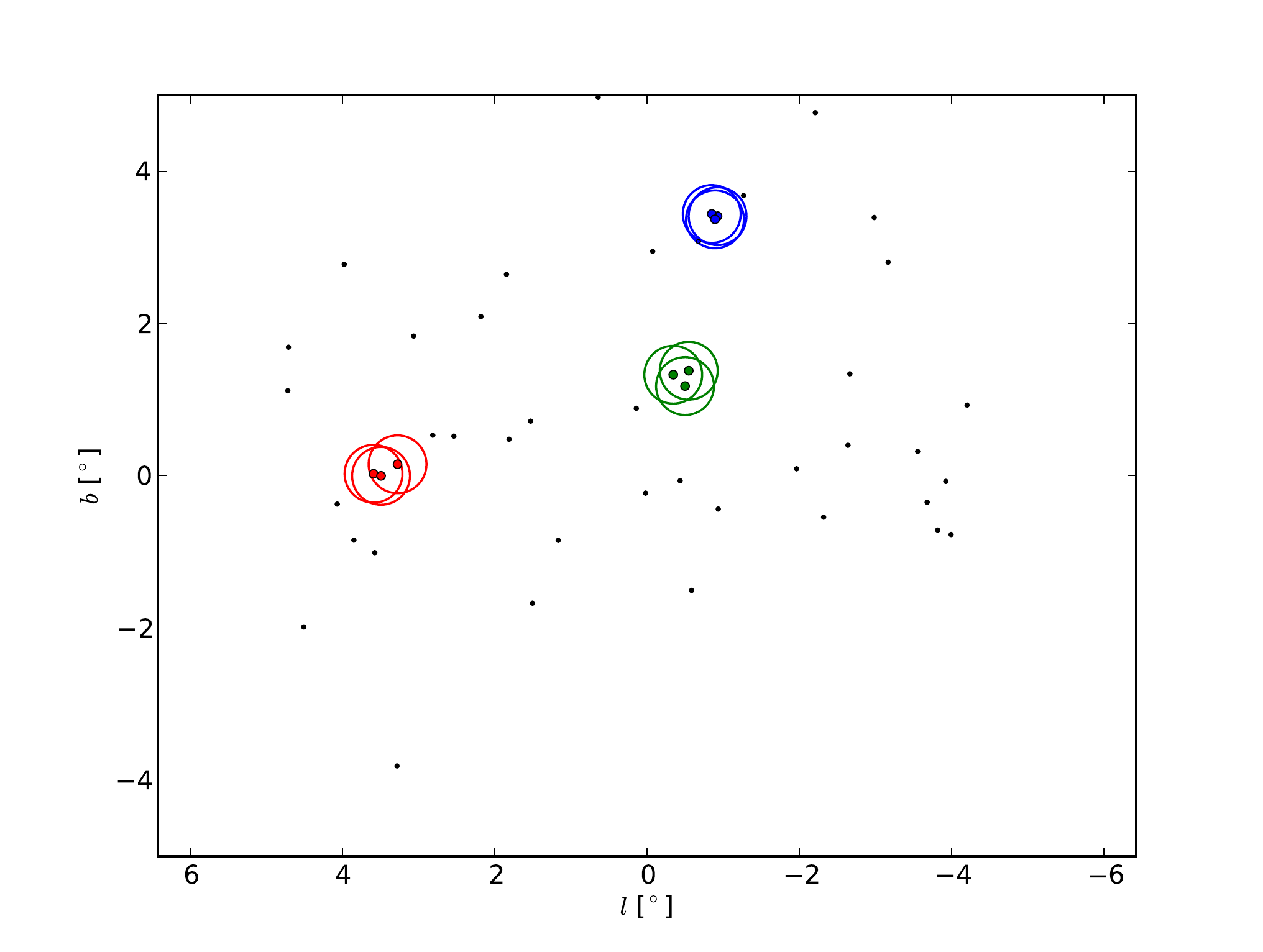}
  \end{tabular}
  \caption{Event map of Fermi photons between 120-140 GeV (left) and a sample
    3 pulsar Monte Carlo simulation (right) showing in colored circles the
    DBSCAN $\epsilon-$neighborhoods for core points in each detected cluster.}
  \label{fig:dbscan}
\end{figure*}

\subsection{Clustering Algorithm}\label{sec:clustering}
In order to classify the spatial morphology of photons in a statistically
robust way, we employ the \textit{Density Based Spatial Clustering of
Applications with Noise} (DBSCAN) algorithm~\citep{dbscan}, which is capable
of both distinguishing cluster points from noise and constraining the
maximum connectivity size based on the instrumental point-spread function.
DBSCAN possesses two input parameters, corresponding to the assumed radius
($\epsilon$) of each cluster neighborhood and the number of points (N$_{\rm
min}$) which must be contained within a neighborhood to form a new cluster or
add to an existing cluster.  Our implementation of DBSCAN, modified from the
Scikit-Learn {\tt python} package \cite{scikit-learn}, works as follows: 

\begin{enumerate}
  \item For each point in the input list, define the `$\epsilon$-neighborhood'
    as a circle of radius $\epsilon$ centered on the point of interest.  
  \item If a point's $\epsilon$-neighborhood contains greater than $N_{\rm
    min}$ points a new cluster is formed and  that point is marked as a `core
    point'.
  \item Two core points are considered `density-connected' if the points are
    mutually contained with in each other's $\epsilon$-neighborhoods.  All
    density connected points are then merged into a single cluster.
  \item Require any cluster to contain at least 3 core points.
\end{enumerate}

Traditional DBSCAN implementations also define the notion of `density
reachable' to indicate points which are not themselves density-connected, but
which lie within a core point's $\epsilon$-neighborhood. This property is,
however, not symmetric, and cluster assignment in general depends on the input
ordering of the data. Our algorithm ignores density-reachable points, thus
ensuring deterministic results. While we assume, in this analysis, that each
profile is centered at the position of the GC, our results are independent of
this assumption as the DBSCAN algorithm focuses on the relative position
between photons, oblivious to any zero point of the profile. We note that a
slightly modified DBSCAN algorithm has already been employed on Fermi-LAT
data in the past~\citep{dbscanfermi}. 

To exemplify the use of the DBSCAN algorithm, Figure \ref{fig:dbscan} shows
the DBSCAN analysis of the Fermi-LAT photon events measured with an energy
between 120 -- 140 GeV (left), and of a simulated model containing two point
sources near the galactic center (right). In each case, we show the DBSCAN
$\epsilon$-neighborhoods for each core point of each detected cluster. For the
Fermi results,
DBSCAN finds only one cluster (interestingly centered on the actual Galactic
center location!), while in the 3 pulsar simulation case, the algorithm
correctly identifies three clusters, at the positions corresponding to where the
pulsar photons were generated. The following discussion explains in detail
the procedure we employ to apply DBSCAN to $\gamma$-ray data.

To quantitatively compare our models against \emph{Fermi} data, we follow
\citet{dbscanfermi} and employ the likelihood ratio proposed by
\citet{likelihood_algorithm} to calculate the cluster significance, $s$, in
terms of the number of cluster photons $N_s$ and background photons $N_b$:
\begin{equation}
  s = \sqrt{ 2 \Bigg( N_{s}\ln \Bigg[ \frac{2N_{s}}{N_{s} + N_{b}} \Bigg] +
  N_{b}\ln \Bigg[\frac{2N_{b}}{N_{s} + N_{b}}\Bigg] \Bigg)}\;.
\end{equation}
Here, $N_{b}$ represents the expected background counts, determined by
integrating a diffuse background model (discussed in Subsection
\ref{subsec:bg_model}), while $N_{s}$ is based on the total photon count
contained in the cluster; we effectively adopt $\alpha=1$ in the notation of
Ref.~\cite{likelihood_algorithm}. According to Ref.
\cite{likelihood_algorithm}, as long as $N_s$ and $N_b$ are not too sparse,
one can equate a cluster with significance $s$ to an ``$s$-standard deviation
observation''.  Thus a cluster significance $s=2$ implies the cluster is a
$2\sigma$ fluctuation above the mean background as computed in Subsection
\ref{subsec:bg_model}. We will use this nomenclature in our analysis.
With the individual cluster significance in hand, we now define the
\textit{``global'' significance} $S$ as the mean significance of each detected
cluster weighted by the number of photons in that cluster.  We then optimize
the choices of the DBSCAN parameters $\epsilon$ and $N_{\rm min}$ for ACT
simulations by maximizing the global significance for the clustering results
from our pulsar simulations.  Finally we explain why this optimization
procedure does not work with the limited Fermi photon count at 130 GeV and
choose appropriate DBSCAN parameters based on the Fermi spatial point spread
function.

\begin{figure*}
  \includegraphics[width=.9\linewidth]{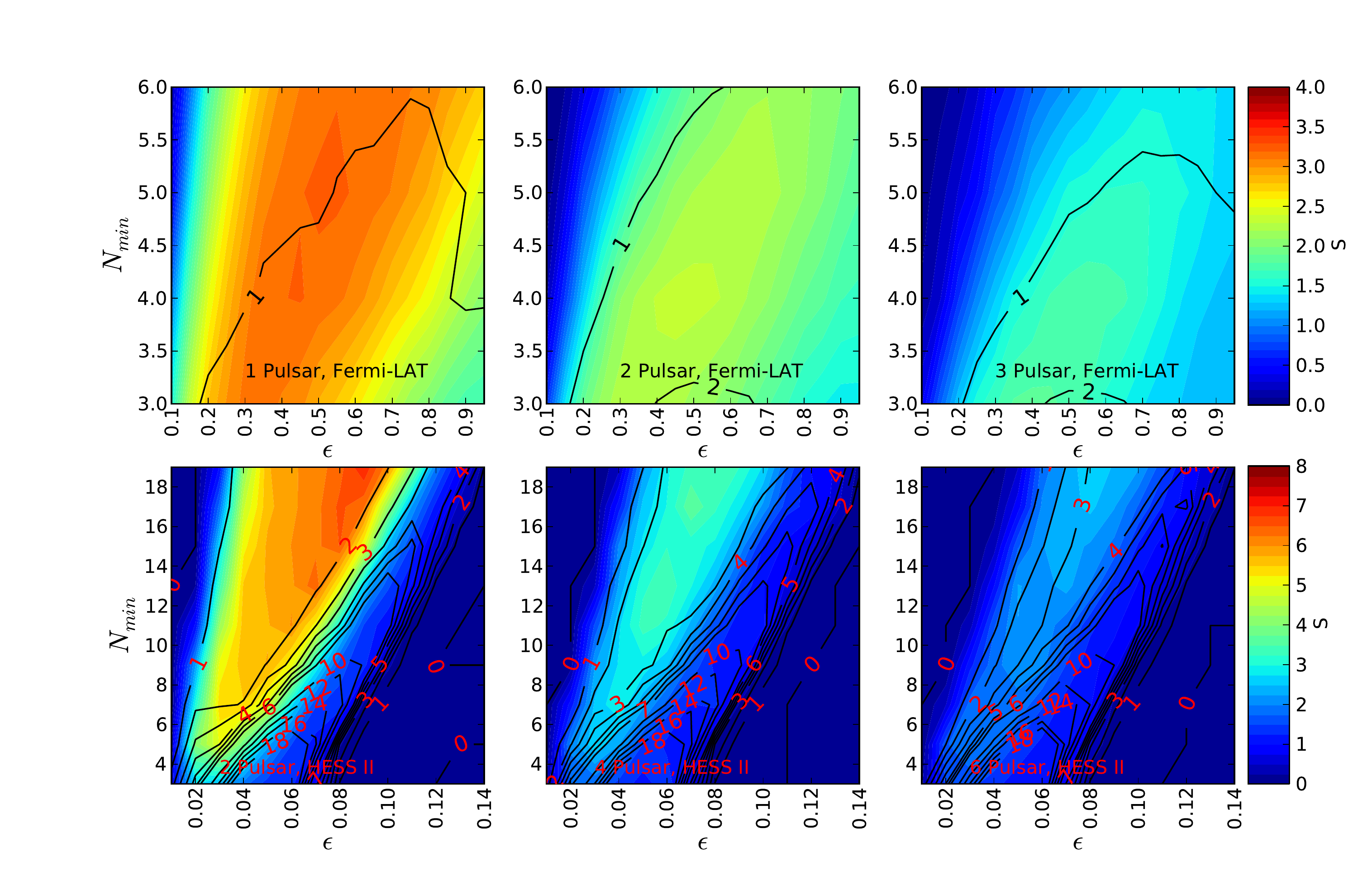}
  \caption{\textit{Global cluster significance} $S$ (filled colored contours)
    and \textit{total number of clusters} $N_{\rm clusters}$ for Fermi (ACT)
    simulations (above threshold $s>1.29$ ($s>2.0$); labeled contours) as a function of
    the DBSCAN search radius $\epsilon$ and core-point threshold $N_{\rm min}$.  The top row corresponds to Fermi simulations of 1 (left),2 (center), and 3 (right) pulsar models while the bottom row is
    for ACT observations of 2 (left), 4 (center), and 6 (right) pulsar models.
    Results are relatively insensitive to large coincident regions in
    the scan parameter space for left and center columns, while the dependence on $N_{\rm min}$ increases
    as the number of photons per pulsar approaches the background rate.  Fermi simulation DBSCAN parameters are chosen to be $\epsilon=0.35$, $N_{\rm
    min}=3$  while ACT
    simulation DBSCAN parameters are chosen to be $\epsilon=0.05$, $N_{\rm
    min}=8$ as a compromise between the cluster detection efficiency and the
    significance over background.  }
  \label{fig:eps_vs_nmin}
\end{figure*}

The value of $\epsilon$ must be large enough that true cluster
elements are not excluded, but small enough that noise is not included. The
variable $\epsilon$, as a result, is closely tied to the physical size of the
instrumental PSF. One must additionally choose a value N$_{\rm min}$ large
enough such that the background does not easily fluctuate above this number,
but low enough that one has a high efficiency of finding real clusters.  To
this end, we use simulations of 1, 2, and 3 pulsar models for Fermi simulations, and 2, 4, and 6 pulsar models for ACT observations to determine a
region of ($\epsilon, N_{\rm min}$) parameter space which simultaneously
optimizes the significance and detection efficiency.  Displayed in the top row of Figure
\ref{fig:eps_vs_nmin} is the global clustering significance (shown in filled contours) and number of detected clusters with $s>1.29$ (inset, labeled contours) for 48 photon Fermi simulations of 1, 2, and 3 pulsars as a function of the DBSCAN parameters
$\epsilon$ and $N_{\rm min}$.  In the bottom row, we again plot the global clustering significance and number of detected clusters with $s>2$  for 5000 photon
ACT simulations with columns from left to right corresponding to 2, 4,
and 6 pulsar models. We note that we apply a firm cut that $N_{\rm min}$ must
be at least 3, as this represents the lowest possible non-trivial clustering
which may be analyzed by the DBSCAN algorithm. 

Inspection of the all three columns for both Fermi and ACT simulations reveals that the
clustering algorithm detects clusters at high significance over large coincident regions
of DBSCAN parameter space. In the case of ACT observations where the clusters are better differentiated from the background, we also see that these regions also detect the correct number of clusters until the number of pulsars becomes to large to reliably detect all true clusters. This indicates that the results are robust for most
reasonable choices of scan parameters while in the case of 6 pulsars, the number of detected clusters is somewhat more sensitive to parameter choices. For these ACT simulations, we see that
choosing $N_{\rm min}$ too small, or $\epsilon$ too large can lead to the
identification of extra (false) clusters which lowers the overall
significance.  

We choose our scan parameters based on the 6 pulsar simulations (bottom right) which
have the lowest signal to noise ratio among the models we consider here.
These considerations motivate a choice of $\epsilon=0.35^\circ$, $N_{\rm
min}=3$ for Fermi simulations and $\epsilon=0.05^\circ$, $N_{\rm
min}=8$ for ACT projections as a balance between preserving significance and detecting most of the clusters at $s>2$.  We note that detailed studies on the
behavior of DBSCAN settings applied to Fermi-LAT data at lower energies have
found qualitatively comparable optimization regions for DBSCAN parameters
~\citep{dbscanfermi}.

In summary, for ACT observations we expect $\sim$5000 photons for a 6h
exposure, and use our significance measure balanced against the number of
detected clusters to optimize the DBSCAN parameters. We find
$\epsilon=0.05^\circ$ and N$_{\rm min}$~=~8.   For Fermi observations, we choose $\epsilon~=~0.35^\circ$ and N$_{\rm min}$~=~3, which represents the lowest
level of non-trivial clustering.  We note that there are only 48 photons in
our sample, and thus we do not expect to be able to identify more than a few
clusters corresponding to point sources with our analysis technique.

\section{Results}
\label{sec:results}
In order to compare our models of the expected 130~GeV line signal produced by
both dark matter and pulsars, we first calculate the clustering properties of
the actual Fermi dataset in the energy range of 120-140~GeV using the DBSCAN
algorithm.  We find only one detected cluster with a significance of $s=1.29$,
an angular scale of 0.22$^\circ$ (defined as the mean pairwise distance of
each pair of cluster members), and 3 member photons (see
Fig.~\ref{fig:dbscan}, left).

\begin{figure*}
  \includegraphics[width=.9\linewidth]{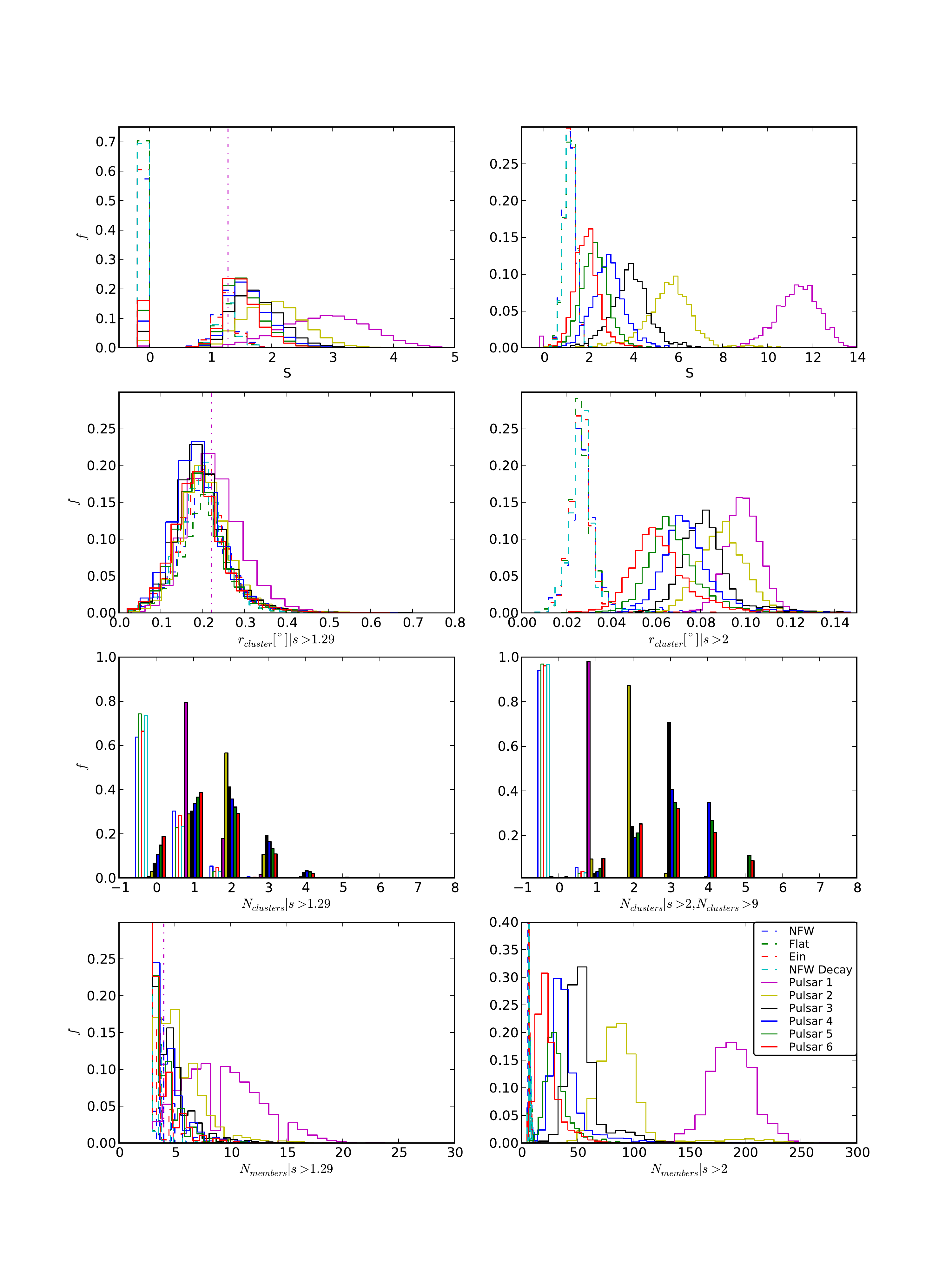}
  \caption{\label{fig:clustering_fermi} Models for the clustering properties
    expected from both Fermi-LAT (48 photons total, left) and H.E.S.S.-II
    (5000 photons total, right) observations of annihilating dark matter
    following a NFW profile (blue dashed), flat density profile (green
    dashed), Einasto profile (red dashed) and decaying dark matter following
    an NFW profile (cyan dashed), as well as models of emission from
    undetected groups of one (magenta solid), two (yellow), three (black),
    four (blue), five (green), and six (red) pulsars, compared to the
    clustering properties observed in the Fermi-LAT data binned from
    120-140~GeV (magenta dot dash). The top row shows the distribution of
    global significance of detected clusters (S, top row).  All other
    quantities are calculated in the subspace of clusters with significance
    $s>1.29$ $(s>2)$ for Fermi (ACT) simulations. Shown is the distribution of
    \textit{mean clustering radii} (2nd row), the distribution of the
    \textit{total number of clusters} detected (3rd row), and the distribution
    of the average \textit{number of member photons} in each cluster with
  $s>2$ (bottom).}
\end{figure*}

We first define the parameters useful for differentiating different emission
classes and then compare to current Fermi data and projections for upcoming
H.E.S.S.-II observations. In addition to
the global significance, $S$, we define three quantities in the space of
clusters with significance $s>$~1.29 for Fermi and $s>$~2 for ACT
observations. The quantities we employ are chosen to best capture the results
of each simulation based on the output of DBSCAN, and provide useful
information on the clustering properties of each source model. Specifically,
we use:
\begin{enumerate}
  \item the \textit{mean clustering radius}, $r_{\rm cluster}$, defined as the
    average of the mean pairwise distance (angular scale) of each cluster
    above threshold, weighted by the cluster's significance;
  \item $N_{\rm clusters}$ defined as the \textit{total number of clusters}
    detected above threshold with at least 3 (10) cluster members for Fermi (ACT), and
  \item $N_{\rm members}$ defined as the \textit{mean number of photons} of
    clusters above threshold, weighted by each cluster's significance.  The
    significance weighting is used simply to suppress the influence of
    clusters which are likely background fluctuations.
\end{enumerate}

In Figure~\ref{fig:clustering_fermi} we show the results of the DBSCAN
algorithm applied to both the Fermi data (vertical dashed-dotted line),
compared to results from Monte Carlo simulations of dark matter and point
source emission from pulsars for 48 photons (left; resembling Fermi-LAT
observations) and 5000 photons (right; resembling ACT observations).  The
dashed lines correspond to diffuse dark matter annihilation models -- NFW
(blue dashed), Einasto (red dashed), NFW decay model (cyan dashed) and a flat
distribution (green dashed). Pulsar models are represented by solid lines -- 1
(magenta), 2 (yellow), 3 (black), 4 (blue), 5 (green), and 6 (red). Each
histogram is normalized and if no clusters are found, the significance
defaults to zero. We thus see that it is much less likely for diffuse models to
produce clusters dense enough to be picked up by DBSCAN, indicating that our
combination of $\epsilon$ and $N_{\rm min}$ are reasonably efficient at
rejecting spurious background clusters, especially in the ACT case.

The \textit{global cluster significance} $S$ (top row of
Figure~\ref{fig:clustering_fermi}) provides the strongest metric for
differentiating point-source from diffuse emission.  Diffuse models should
possess virtually identical clustering properties as the extent of any true
structure is much larger than the instrumental point spread function.
Thus, diffuse sources should only occasionally produce low significance,
loosely grouped clusters due to background fluctuations. One possible
exception for dark matter models is the identification of a single point
source at the GC where the dark matter annihilation rate can be very large.
It is notable that the single cluster found in the Fermi data lies precisely
on the galactic center.  \EC{We see in the case of Fermi simulations (top left
panel) that there is a fairly sharp cutoff in the fraction of models with
global significance $S\lesssim1$.  This lower bound is set by DBSCAN's minimum
cluster detection requirements, as well as the location of the cluster with
respect to the background template, which determines the number of background
photons in that region.  Because the cluster detected in the Fermi data is
loose, we expect the single detected Fermi cluster to be close to the
effective cutoff ($S = 1.29$ for the detected cluster).}

\begin{table}[t]
  \centering
  \begin{tabular}{cccccccc}
    \toprule 
   Model   & 
   $f_{s>1.29}^{\textsc{\tiny{Fermi-LAT}}}$  &   $f_{s>1.29}^{\textsc{\tiny{Fermi-LAT}}}$ &
   $f_{s>2.0}^{\textsc{\tiny{ACT}}}$ & $f_{s>2.0}^{\textsc{\tiny{ACT}}}$ & $f_{s>2.0}^{\textsc{\tiny{ACT}}}$
   \\ & $n=1$ & $n=2$ &$n=1$&$n=2$&$n=3$\\\hline\hline
	NFW      & 0.362  & 0.059 & 0.060 & 0.002 & 0.0\\     
	Einasto  & 0.335  & 0.052 & 0.039 & 0.001 & 0.0\\     
	NFW Decay& 0.264  & 0.031 & 0.034 & 0.001 & 0.0\\     
   	Flat     & 0.257  & 0.030 & 0.031 & 0.001 &  0.0\\     
    1 Pulsar  & 0.992 & 0.197 & 0.984    & 0.003 &  0.0\\
    2 Pulsars & 0.971 & 0.681 & 0.997    & 0.902 & 0.030\\
    3 Pulsars & 0.934 & 0.631 & 0.998    & 0.967 & 0.726\\
    4 Pulsars & 0.897 & 0.556 & 0.997    & 0.956 & 0.769\\
    5 Pulsars & 0.851 & 0.485 & 0.993    & 0.942 & 0.730\\
    6 Pulsars & 0.811 & 0.424 & 0.985    & 0.888 & 0.635\\
    \hline
  \end{tabular}
  \caption{The fraction of simulations with at least 1 cluster of
  significance $s>1.29$ ($n=1$ and $s=1.29$, corresponding to the maximum $n$
  and $s_i$ for clusters $i$ found in Fermi-LAT data) for Fermi-LAT
  simulations (column 2), the of Fermi-LAT simulations with at least 2 clusters of significance $s>1.29$ (column 3) and fraction of simulations with at least $n$
  clusters detected at significance $s>2.0$ with at least 10 core points for ACT simulations (columns
  4-6).}
  \label{tab:pulsars}
\end{table}

The second row of Figure~\ref{fig:clustering_fermi} shows the \textit{mean
clustering radius}.  For ACT observations there is a
clear division between the point source and
smaller diffuse emission scales. However, the Fermi-LAT
simulations do not offer any useful discrimination between models with such low photon counts. For point
sources, this distribution is governed by the average value of the PSF and
possesses an asymmetric tail at larger scales due to the inclusion of
background photons and events whose true position is determined by the long
PSF tails.  For diffuse models, the distribution is governed dominantly by the
$\epsilon$ DBSCAN parameter until the background density becomes dominant.

Displayed in the third row is the distribution of the \textit{total number of
clusters}, $N_{\rm clusters}$, found with significance s~$>$~1.29 (s~$>$~2).  In the case of ACT observations we also require clusters to have at least 10 core points to be included whereas we required only 3 for Fermi.
We expect to be able to discern \emph{at most} $N_{sig}/N_{\rm min}$ point
sources if the signal photons happen to distribute themselves evenly between
sources.  Even in this case, these true clusters may still lie below the
significance threshold.  It is realistic to identify only 2-3 true clusters
with 12 signal photons (48 total) if they in fact have point source
progenitors.  This is reflected here.  For diffuse sources, only a fraction the
detected clusters pass the significance cut. As the number of events is
increased, this significance cut could be increased to maintain a high
acceptance/rejection ratio, though it is clear that for diffuse models, we
typically obtain either zero clusters or one cluster per simulation with our
current significance cuts, while still efficiently detecting several clusters
in the case of pulsar models.

Finally, the fourth row contains the distribution of the \textit{mean number
of cluster members}, $N_{\rm members}$.  For a random distribution of events
between pulsars, we expect a Poissonian distribution with a mean of
approximately $N_{sig}/N_{pulsars}$ (contributions from the background should
typically be $<$1 photon for Fermi, but the number of signal photons can
fluctuate significantly during Poisson sampling).  For Fermi-LAT observations
we expect $12\pm 3.5$ signal photons distributed between $N_{pulsars}$.  In
the case of ACT observations, we expect $232\pm 15$ photons. Occasional
spurious clusters will force this distribution downwards, although this effect
is reduced by the significance weighting and because we only consider ``core
points'' to be cluster members, thus rejecting those lying on the cluster
boundaries.  \bigskip

In order to quantify in how confidently point-source or diffuse models for the 130 GeV excess
can be rejected, we count the fraction of simulations which are incompatible
with Fermi-LAT data for each tentative source class. A simulation is deemed
incompatible if at least 1 cluster is detected at significance $s>1.29$
corresponding to the maximum number and significance of clusters detected in
the Fermi-LAT data.  The first column of Table \ref{tab:pulsars} shows the
fraction of simulations for each model with at least one cluster ($n=1$) with
$s_i>1.29$. In column 3 we show the fraction of simulations which have two
clusters detected with a significance of $s_i>1.29$ in Fermi-LAT simulations,
which further demonstrates the vast statistical separations between the
clustering properties of diffuse and point-source models. In the subsequent
columns we show similar data for the H.E.S.S.-II telescope, which also requires a cluster to have at least 10 core points, and clearly provides an
ever greater ability to differentiate between source classes producing the
$\gamma$-ray line. 

Our statistical approach shows that Fermi-LAT data already rule out models
where the 130 GeV $\gamma$-ray line is produced by 1, (2, 4) or fewer pulsars at the 99\%
(95\%, 90\%) confidence level (CL). Specifically, only one cluster was detected in the
Fermi-LAT data with a significance $s=1.29$, while a cluster with a larger
significance is observed in more than  90\% of simulations with any ensemble
of less than 4 point sources. Due to the greatly increased effective area of
the H.E.S.S.-II telescope, we find an even greater statistical separation
between our models of diffuse and point source emission. This indicates that
H.E.S.S.-II will be able to conclusively differentiate models of the 130 GeV
line using purely statistical properties. 

\medskip

\section{Discussion and Conclusions}
\label{sec:conclusions}
If confirmed, the tentative detection of a $\gamma$-ray line in the Fermi data
might potentially turn into one of the most important breakthroughs on physics
beyond the Standard Model, pointing towards the mass of the dark matter
particle. Key future developments include the analysis of Fermi $\gamma$-ray
events with the forthcoming Pass 8 version of the Fermi-LAT analysis software,
which will include a major overhaul of the energy reconstruction algorithm
\cite{pass8}. It will also be crucial to identify whether the excess events at
130 GeV from the Earth's limb are indeed a statistical fluke. This question
will be answered by increased exposure, which will increase the current
statistical sample \cite{su_finkbeiner_line, bloom_charles_fermi_lat_line}.

At present, barring instrumental effects, a 130 GeV line could be ascribed to
either new physics, presumably a dark matter particle decaying or
pair-annihilating into a $2\gamma$ (or $\gamma$Z, $\gamma$h etc.) final state, or to
one or more pulsars featuring a cold wind with electrons with an energy at, or
very close to, 130 GeV. This latter possibility, it is argued in
\citet{aharonian_cold_pulsar_winds}, is the only ``traditional'' astrophysical
process envisioned thus far that could produce a sharp gamma-ray line in the
energy regime of interest. It is therefore of the utmost importance to
discriminate between a cold pulsar wind scenario and a dark matter scenario,
if indeed the line is resilient to future tests and observations.

Discriminating dark matter and pulsar interpretations of the 130 GeV line may
not be possible based solely on the spectral characteristics of the Fermi-LAT
data. In the present study we
sought to use morphological information, i.e. the 130 GeV events' arrival
direction, to establish whether the signal is likely due to multiple point
sources as opposed to a truly diffuse origin. This is a meaningful question,
since the signal region is much larger than the instrumental angular
resolution, and it should be thus possible to discriminate a finite number of
point sources, as expected in the pulsar case, from a distribution that
follows a diffuse morphology, such as what expected from dark matter
annihilation or decay.

To quantitatively approach the issue of discriminating pulsars versus dark
matter on a morphological basis, we employed the DBSCAN algorithm which
distinguishes clusters from background noise based on the local photon
density. We defined a statistical significance measure, and we optimized the
algorithm's two physically well-constrained parameters in order to reconstruct
as accurately as possible the potential ``clusters'' producing the observed
$\gamma$-ray events.

As a result of our analysis of the available Fermi-LAT data, we concluded that
at the 99\%, 95\%, 90\% confidence level the data need at least 2, 3, 5 or more point sources, respectively,
while the events' morphology is perfectly consistent with various dark matter
density profiles.  We conclude that the data strongly disfavor the hypothesis
of a small number of pulsars as the origin of the signal. If the pulsar
scenario is indeed the culprit for the 130 GeV events, it is necessary to
postulate a relatively large population  of pulsars  {(likely at least 4)}
with cold winds featuring electrons with exactly, to within the instrumental
energy resolution, the same energy. This appears, to say the least, quite
problematic.  A diffuse origin seems therefore the most likely scenario for
the 130 GeV photons. Our clustering algorithm approach, clearly, is not
optimized to discriminate between different diffuse morphologies: in fact, on
the basis of our results, we find that we cannot discriminate between
different diffuse morphologies (like dark matter annihliation vs.~decay).

Present and future observatories have the potential to shed additional light
on the presence and characteristics of the 130~GeV
line~\citep{bergstrom_future_observatories}. Improvements to the H.E.S.S.
telescope (H.E.S.S.-II) have reduced the $\gamma$-ray threshold to around
50~GeV, allowing for the independent determination of a line signal from the
GC region. This is especially important, as the 10$^4$~m$^2$ collecting area
of the H.E.S.S. telescope will quickly alleviate the low-statistics issues
involved in Fermi-LAT studies~\citep{aharonian_cold_pulsar_winds}.
Furthermore, future instruments such as Gamma-400 \citep{gamma_400_status} and
CTA~\citep{cta_status} are likely to provide the necessary effective area and
energy-resolution to definitively and conclusively test the existence and
nature the 130~GeV line feature. 

However, in the near future, the most important contribution is expected to
come from Fermi-LAT itself: Additional data taken since last year, and the
continuous accumulation of more data over the next years, will show whether
the signature persists or is a rare statistical fluke. Simultaneously, the
availability of pass 8 events, based on a set of completely rewritten event
reconstruction algorithms for the LAT, will allow a fresh look on possible
instrumental systematics.

\acknowledgments
This work is partly supported by NASA grant NNX11AQ10G. SP also acknowledges
partial support from the Department of Energy under contract
DE-FG02-04ER41286. \\

\bibliography{line} 

\end{document}